\documentclass[twocolumn]{aastex62}
\shorttitle{Short period planets as possible strange quark objects}
\shortauthors{Abudushataer Kuerban et al.}

\begin{document}

\title{Close-in Exoplanets as Candidates of Strange Quark Matter Objects}
\correspondingauthor{Y. F. Huang; J. J. Geng; A. Kuerban;}
\email{hyf@nju.edu.cn; gengjinjun@nju.edu.cn; lompa46@163.com}

\author[0000-0002-0786-7307]{Abudushataer Kuerban}
\affil{School of Astronomy and Space Science, Nanjing University, Nanjing 210023, China}

\affiliation{Key Laboratory of Modern Astronomy and Astrophysics (Nanjing University),
	Ministry of Education,  Nanjing 210023, China}

\author{Jin-Jun Geng}
\affil{School of Astronomy and Space Science, Nanjing University, Nanjing 210023, China}

\affiliation{Key Laboratory of Modern Astronomy and Astrophysics (Nanjing University),
	Ministry of Education, Nanjing 210023, China}

\author{Yong-Feng Huang}
\affil{School of Astronomy and Space Science, Nanjing University, Nanjing 210023, China}
\affiliation{Key Laboratory of Modern Astronomy and Astrophysics (Nanjing University),
	Ministry of Education, Nanjing 210023, China}

\author{Hong-Shi Zong}
\affil{Department of Physics, Nanjing University, Nanjing 210093, China}
\affiliation{Joint Center for Particle, Nuclear Physics and Cosmology, Nanjing 210093, China}
\affiliation{State Key Laboratory of Theoretical Physics, Institute of Theoretical Physics, CAS, Beijing 100190, China}

\author{Hang Gong}
\affil{National Astronomical Observatories, Chinese Academy of Sciences, Beijing 100101, China}




\begin{abstract}

Since the true ground state of the hadrons may be strange quark matter (SQM),
pulsars may actually be strange stars rather than neutron stars.
According to this SQM hypothesis, strange planets can also stably exist.
The density of normal matter planets can hardly be higher than 30~g~cm$^{-3}$.
As a result, they will be tidally disrupted when its orbital radius is less than
$\sim 5.6\times10^{10} \rm \, cm $, or when the orbital period ($P_{\rm orb}$) is less
than $ \sim \rm 6100\, s $. On the contrary, a strange planet can safely
survive even when it is very close to the host, due to its high density.
The feature can help us identify SQM objects.
In this study, we have tried to search for SQM objects among close-in exoplanets
orbiting around pulsars. Encouragingly, it is found that four pulsar planets
(XTE J1807-294 b, XTE J1751-305 b, PSR 0636 b, PSR J1807-2459A b) completely
meet the criteria of $P_{\rm orb} < \rm 6100\, s $, and are thus good candidates
for SQM planets. The orbital periods of two other planets (PSR J1719+14 b and
PSR J2051-0827 b) are only slightly higher than the criteria. They could
be regarded as potential candidates. Additionally, we find that the periods of
five white dwarf planets (GP Com b, V396 Hya b, J1433 b, WD 0137-349 b, and SDSS J1411+2009 b)
are less than 0.1 days. We argue that they might also be SQM planets.
It is further found that the persistent gravitational wave emissions from at least
three of these close-in planetary systems are detectable to LISA. More encouragingly,
the advanced LIGO and Einstein Telescope are able to detect the gravitational wave
bursts produced by the merger events of such SQM planetary systems,
which will provide a unique test for the SQM hypothesis.

\end{abstract}



\section{Introduction} \label{sec:intro}

Soon after the discovery of neutrons, the existence of neutron stars (NS), which are mainly
made up of neutrons, was predicted. In 1960's, pulsars were discovered. As extremely compact
objects with a typical mass of $\sim 1.4M_\odot $ and a typical radius of only about $ 10\rm\, km $,
they were soon identified as neutron stars. However, it has also been argued that the true ground
state of the matter at extreme densities may actually be quarks \citep{Itoh1970,Bodmer1971} rather than the hadronic form.
The internal composition of these extremely compact stars thus is still largely unclear.
For instance, under such an extreme condition, some particles like hyperons, baryons, and even
bosons may appear; quark deconfinement may also happen. In particular, it has long been suggested
that even more exotic states such as strange quark matter (SQM) may exist in the core \citep{Itoh1970,Bodmer1971,Witten1984,Farhi1984}.
Recently, the discovery of several $ 2M_\odot $ pulsars \citep{Demorest2010,Antoniadis2013,Cromartie2019} attracts the attention
of scientists. Pulsars with such a high mass and a small radius imply that the density at the center
can reach several times of nuclear saturation density, which further complicates the internal composition
of these compact stars.

Following the SQM hypothesis, the existence of a whole sequence of SQM objects, such as strange quark
stars (SSs) \citep{Witten1984,Farhi1984,Alcock1986}, strange quark dwarfs \citep{Glendenning1995,Glendenning1995+}, and strange quark
planets \citep{Glendenning1995,Glendenning1995+,Xu2003,Horvath2012,Huang2017} are predicted. For example, \citet{Jiang2018} argued that
the double white dwarf binary J125733.63+542850.5 may actually contains two strange dwarfs.
SQM objects may be covered by a thin
crust of normal hadronic matter, or may even simply be bare SQM cores \citep{Glendenning1995,Glendenning1995+}. The common
compact nature of SSs and NSs makes it difficult to discriminate these two kinds of internally different
stars observationally \citep{Alcock1986}. A few efforts have been made to reveal the difference between them.
For example, they may have different $ M-R $ relations \citep{Witten1984,Krivoruchenko1991,Glendenning1995,Li1995,Avellar2010,Drago2014},
and SSs may rotate much faster (with spin period $ P_{\rm spin}<1\rm ms $) than
NSs \citep{Friedman1989,Frieman1989,Glendenning1989,Kristian1989,Madsen1998,Dai1995,Dai1995+,Sawyer1989,Bhattacharyya2016}. They may also have different cooling
rates \citep{Pizzochero1991,Page1992,Ma2002}, different gravitational wave (GW)
features \citep{Jaranowski1998,Madsen1998,Lindblom2000,Andersson2002,Jones2002,Bauswein2010,Moraes2014,Geng2015,Mannarelli2015},
different maximum masses  \citep{Lai2009,Li2010,Weissenborn2011,Mallick2013,Zhu2013,Zhou2018,Shibata2019}, and so on.
Nevertheless, due to the impracticability of the above methods at the current stage, the
problem still remains unsolved.

Encouragingly, several new methods were recently proposed to distinguish SSs from NSs.
The basic idea involves the tremendous difference between SQM planets and normal matter ones.
Because of the extreme compactness, an SQM planet can be very close to its host SS star,
without being tidally disrupted. It can even emit strong GW signals when it finally spirals-in and merges
with the host star\citep{Geng2015}. GW emission from these merging SQM planets within our Galaxy can be
detected by GW detectors such as advanced LIGO and the future Einstein Telescope.
It is thus suggested that we could identify SQM objects by searching for very
close-in planets around pulsars \citep{Huang2017}, or by detecting GW bursts from merging SQM planet
systems \citep{Geng2015}.

It is interesting to note that nearly ten GW events from merging double black holes (and even one
from merging double neutron stars) have been detected by advance LIGO and Virgo since
2016 \citep{Abbott2016,Abbott2017}.
Recently, advanced LIGO has just begun a new observational run, which will surely come up with much more
GW events. The great breakthrough in GW astronomy hopefully sheds light on possible detection
of GW emission from merging SQM planet systems in the near future.
At the same time, rapid progress in observational technology also leads to a drastic increase
in the number of extrasolar planets being detected in the past decades. Interestingly, a good number
of exoplanets are found to be orbiting around pulsars. In this study, we examine these pulsar
planets systematically to search for very close-in ones that could be ideal candidates for SQM
objects. The possibility of detecting GW emission from these candidates will also be explored.

The structure of our paper is organized as follows. In Section \ref{sec:Theories}, the background
relevant to SQM planet systems is briefly introduced. In Section \ref{sec:data}, we describe the data
source of our sample. In Section \ref{sec:close-in}, SQM candidates are selected and evaluated by
considering the criteria of close-in introduced in Section \ref{sec:Theories}. GW emission from
the candidate SQM planet systems is calculated and compared with the limiting sensitivities of
current and future GW experiments in Section \ref{sec:GW}. Finally, Section \ref{sec:conclusion}
presents our conclusions and discussion.

\section{Theories relevant to SQM planet systems} \label{sec:Theories}

\subsection{Criteria for identifying SQM planets} \label{subsec:criteria}

\begin{deluxetable*}{lccclccc}[t!]
	\tablecaption{Candidate pulsar planets and their host pulsars. \label{tab:table1}}
	\tablewidth{0pt}
	\tablehead{
		\colhead{Planet name} & \colhead{Mass}  & \colhead{$ P_{\rm orb} $}  &\colhead{Host name} & \colhead{Distance} & \colhead{Mass} & \colhead{$ Ref.$}\\				
		\colhead{} & \colhead{$ m $ ($ M_{\rm jup} $)} & \colhead{(day)} & \colhead{} & \colhead{$ d $ ($\rm pc $)} & \colhead{$ M $ ($ M_\odot $)} &
	}
	\startdata
	\multicolumn{7}{c}{Gold sample}\\
	\cline{1-7}
	PSR 0636 b & 8    & 0.067 & PSR J0636 & 210   & 1.4   & 1, 2, 3 \\
	PSR J1807-2459A b & 9.4  & 0.07  & PSR J1807-2459A & 2790  & 1.4   & 4, 5, 6, 7 \\
	PSR 1719-14 b & 1    & 0.090706293 & PSR 1719-14 & 1200  & 1.4   & 3, 8, 9 \\	
	PSR J2322-2650 b  & 0.7949  & 0.322963997 & PSR J2322-2650  & 230   & 1.4   & 3 \\
	PSR 1257+12 b & 0.00007  & 25.262 & PSR 1257+12 & 710   & 1.4   & 10, 11, 12, 13, 14 \\
	PSR 1257+12 c & 0.013 &  66.5419 & PSR 1257+12 & 710   & 1.4   & 10, 11, 12, 13, 14 \\
	PSR 1257+12 d & 0.012 & 98.2114 & PSR 1257+12 & 710   & 1.4   & 10, 11, 12, 13, 14 \\
	PSR B0943+10 b & 2.8  & 730   & PSR B0943+10 & 890   & 1.5   & 15 \\
	PSR B0943+10 c & 2.6  & 1460  & PSR B0943+10 & 890   & 1.5   & 15 \\
	PSR B0329+54 b & 0.0062 & 10139.34 & PSR B0329+54 & 1000  & 1.4   & 16, 17, 18 \\
	PSR B1620-26(AB) b & 2.5  & 36525 & PSR B1620-26(AB) & 3800  & 1.35  & 19, 20, 21, 22, 23 \\		
	\cline{1-7}
	\multicolumn{7}{c}{Silver sample}\\
	\cline{1-7}
	PSR J2051-0827 b & 28.3  & 0.099110266 & PSR J2051-0827 & 1280  & 1.4   & 7, 24 \\	
	PSR J2241-5236 b & 12    & 0.14567224 & PSR J2241-5236 & 500   & 1.35  & 7, 25 \\
	PSR B1957+20 b & 22    & 0.38  & PSR B1957+20 & 1530  & 1.4   & 7, 26 \\	
	\cline{1-7}
	\multicolumn{7}{c}{Copper sample}\\
	\cline{1-7}        	
	XTE J1807-294 b & 14.5 & 0.0278292 & XTE J1807-294 & 5500  & 1.5   & 27, 28, 29, 30, 31, 32 \\
	XTE J1751-305 b & 27   & 0.02945997 & XTE J1751-305 & 11000 & 1.7   & 33, 34, 35 \\
	PSR J1544+4937 b & 18    & 0.12077299 & PSR J1544+4937 & 3500  & 1.7   & 36, 37 \\
	PSR J1446-4701 b & 23    & 0.277666077 & PSR J1446-4701 & 1500  & 1.4   & 38, 39, 40 \\
	PSR J1502-6752 b & 26    & 2.48445723 & PSR J1502-6752 & 4200  & 1.4   & 38, 39 \\
	\enddata
	\tablecomments{$ Ref.\, $:
		(1) \citet{Stovall2014};
		(2) \citet{Spiewak2016};
		(3) \citet{Spiewak2018};		
		(4) \citet{Amico2001};
		(5) \citet{Ransom2001};
		(6) \citet{Lynch2012};
		(7) \citet{Ray2017};				
		(8) \citet{Bailes2011};
		(9) \citet{Martin2016};		
		(10) \citet{Wolszczan1992};
		(11) \citet{Wolszczan1994};
		(12) \citet{Wolszczan2012};		
		(13) \citet{Patruno2017};
		(14) \citet{Wolszczan2018};				
		(15) \citet{Suleymanova2014};		
		(16) \citet{Demianski1979};
		(17) \citet{Shabanova1995};
		(18) \citet{Starovoit2017};
		(19) \citet{Thorsett1993};
		(20) \citet{Lewis2008};		
		(21) \citet{Mottez2011};		
		(22) \citet{Schneider2011};
		(23) \citet{Veras2016};
		(24) \citet{STAPPERS1996};
		(25) \citet{Keith2011};
		(26) \citet{Reynolds2007};		
		(27) \citet{Markwardt2003a,Markwardt2003b};
		(28) \citet{Campana2003};
		(29) \citet{Kirsch2004};
		(30) \citet{Falanga2005};
		(31) \citet{Riggio2007};
		(32) \citet{Patruno2010};
		(33) \citet{Markwardt2002};
		(34) \citet{Gierli2005}; 
		(35) \citet{Andersson2014};							
		(36) \citet{Bhattacharyya2013};
		(37) \citet{Tang2014};		
		(38) \citet{Keith2012};
		(39) \citet{Ng2014};		
		(40) \citet{Arumugasamy2015}.				
	}	
\end{deluxetable*}

\begin{deluxetable*}{lccclccc}[t!]
	\tablecaption{White dwarf planets with $ P_{\rm orb}<0.1 $ days and
		their host stars. \label{tab:table2}}
	\tablewidth{0pt}
	\tablehead{
		\colhead{Planet name} & \colhead{Mass}  & \colhead{$ P_{\rm orb} $}  &\colhead{Host name} & \colhead{Distance} & \colhead{Mass} & \colhead{$ Ref.$}\\				
		\colhead{} & \colhead{$ m $ ($ M_{\rm jup} $)}  & \colhead{(day)} & \colhead{} & \colhead{$ d $ ($\rm pc $)} & \colhead{$ M $ ($ M_\odot $)} &
	}
	\startdata
	GP Com b & 26.2  & 0.032 & GP Com & 75    & 0.33  & 1, 2, 3, 4 \\
	V396 Hya b & 18.3  & 0.045 & V396 Hya  & 77    & 0.32  & 2, 3, 4, 5 \\
	J1433 b & 57.1  & 0.054 & J1433 & 226   & 0.8   & 3, 4, 6, 7 \\
	WD 0137-349 b & 56    & 0.07943002 & WD 0137-349 & 102.26 & 0.39  & 8, 9, 10, 11 \\
	SDSS J1411+2009 b & 50   & 0.0854 & SDSS J1411+2009 & 177   & 0.53  & 12, 13, 14 \\
	\enddata
	\tablecomments{$ Ref.\, $:
		(1) \citet{Nather1981};
		(2) \citet{Kupfer2016};
		(3) \citet{Wong2018};
		(4) \citet{Cunha2018};		
		(5) \citet{RUIZ2001};
		(6) \citet{Littlefair2006};
		(7) \citet{Santisteban2016};		
		(8) \citet{Maxted2006};
		(9) \citet{Burleigh2006};
		(10) \citet{Casewell2015};
		(11) \citet{Longstaff2017};		
		(12) \citet{DRAKE2010};
		(13) \citet{Beuermann2013};
		(14) \citet{Littlefair2014}.
	}	
\end{deluxetable*}

The tidal disruption radius of a planet by its host star is mainly dependent on the density ($ \rho $)
of the planet and the mass of the central host star ($ M $). It can be expressed
as $r_{\rm td}\approx\left(\frac{6M}{\pi\rho}\right)^{1/3}$ \citep{Hills1975}.
If the planet is an SQM one, which typically has an extremely high density
of $\sim \rm 4\times10^{14}\,g\,cm^{-3}$, then the tidal disruption radius can be
estimated as
\begin{eqnarray}
r_{\rm td} \approx2.37\times10^6\left(\frac{M}{1.4M^{\odot}}\right)^{1/3}
\nonumber \\\times\left( \frac{\rho}{4\times10^{14}\rm \,g\,cm^{-3}}\right) ^{-1/3}\rm cm .
\end{eqnarray}
Taking the host star mass as $ 1.4M_\odot $, the above equation tells us that an SQM planet
with a density of $\rm 4\times10^{14}\,g\,cm^{-3}$ will be disrupted only when its orbital radius
is less than $ 2.37\times10^6 $ cm, i.e. when it almost comes to the surface of the host pulsar.

On the contrary, normal planets typically have a density of 1~---~10 $\rm g\,cm^{-3} $.
If we take $\rm 30\,g\,cm^{-3}$ as an upper limit for the density of normal planets, then the
limiting disruption radius is $ 5.6\times10^{10} $ cm \citep{Huang2017}. In this study, we take
this value as a criteria to discriminate normal planets and SQM ones. If a planet is observed to
have an orbital radius ($a$) smaller than $5.6\times10^{10}\rm\, cm$, then it is most likely an
exotic strange quark object, but not a normal matter planet. According to the Kepler's law,
such a close-in planet should also have a very small orbital
period, $ P_{\rm orb}\lesssim 6100\rm\,s $ \citep{Huang2017}. Therefore, we could identify candidates of
SQM planets by using the criteria of $P_{\rm orb} \lesssim 6100\rm \,s $
and/or $a \lesssim 5.6\times10^{10} $cm.

\subsection{GWs from SQM planet systems}

According to general relativity, orbital motion of a binary system can lead to
GW emission and spiral-in of the system. The GW emission power of a system with known
masses and orbital parameters is,
\begin{equation}
L_{\rm GW}=\frac{32G^{4}}{5c^5}\frac{M^{2}m^{2}\left( M+m\right) }{a^{5}}f\left( e\right),
\end{equation}
where $c$ is the speed of light, $ G $ is the gravitational constant, and $ m $ is the mass
of the planet. The factor $F\left( e\right) =\left(1+\frac{73}{24}e^{2}+\frac{37}{96}e^{4}\right)/\left( 1-e^{2}\right)^{7/2}$
is a function of the orbital eccentricity ($e$). Here we take $F\left( e\right) = 1$ for circular orbits
considered in our modeling.

In a binary system, the orbit will evolve with time due to continuous GW emission. During
this process, the GW strain will increase with time. If the distance of the binary system with
respect to us is $d$, then the strain amplitude of the GW can be expressed as \citep{Peters1963,Postnov2014,Geng2015},
\begin{equation}
h=5.1\times10^{-23}\left( \frac{M_{C}}{1\,M_\odot}\right)^{5/3}
\left( \frac{P_{\rm orb}}{1 {\rm\, hr}}\right)^{ -2/3}\left( \frac{d}{10 {\rm \,kpc}}\right) ^{-1},
\end{equation}
where $ M_{C}=\left( Mm\right)^{3/5}/\left( M+m\right)^{1/5} $ is the chirp mass.
Directly observable quantity of GW is the strain spectral amplitude.
For a binary system, it is given as \citep{Finn1993,Nissanke2010,Postnov2014,Geng2015},
\begin{eqnarray}
h_{f}=6.4\times10^{-21}\left(\frac{M_{C}}{1M_\odot}\right)^{5/6}\left( \frac{f}{300Hz}\right) ^{-7/6}\nonumber \\
\times\left( \frac{d}{10\rm\,kpc}\right)^{-1} Hz^{-1/2},
\end{eqnarray}
where $f$ is the GW frequency that may evolve with time, $f={2}/{P_{\rm orb}}$.

Because of the continuous energy loss through GW emission, the system will coalesce at the
final stage of the inspiraling process. The coalescence time scale \citep{Peters1963, Peters1964, Lorimer2008}
of the system is expressed as
\begin{equation}
{t_{\rm co}}=9.88\times10^{6}{\rm yr}\left(\frac{P_{\rm orb}}{\rm 1\, hr}
\right)^{8/3}\left(\frac{\mu}{1\, M_{\odot}} \right)^{-1}
\left(\frac{\mathcal{M}}{1\,M_{\odot}} \right)^{-2/3},
\end{equation}
where $ \mu={Mm}/\left( M+m\right)$ is the reduced mass,
$ \mathcal{M}=M+m $ is the total mass of the system.

\section{Data collection}\label{sec:data}

In this study, we will systematically examine all the available short period exoplanets
to search for possible candidate SQM planets. For this purpose, we have searched through various
exoplanet data bases. Currently, popular exoplanet data bases that are widely used in the field
include: the Extrasolar Planets Encyclopaedia (hereafter, EU\footnote{\url{http://www.exoplanet.eu/}}),
the NASA Exoplanet Archive (ARCHIVE\footnote{\url{https://exoplanetarchive.ipac.caltech.edu/}}),
the Open Exoplanet Catalogue
(OPEN\footnote{\url{http://www.openexoplanetcatalogue.com/}}), the Exoplanet Data Explorer
(ORG\footnote{\url{http://www.exoplanets.org/}}),
and Extrasolar planet's catalogue produced by Kyoto University
(EXOKyoto\footnote{\url{http://www.exoplanetkyoto.org/catalog/?lang=en}}).
The numbers of planets in these databases are very different from each other. Interestingly,
note that a detailed comparison of these databases has been carried out by \citet{Bashi2018}.
Generally speaking, EU seems to provide the most complete sample for exoplanets.
There are totally 6699 planets listed on the EU web site, among which 4011 are confirmed and 2688 are candidates.

Since SQM planets are most likely to be found orbiting around pulsars (in this case, the pulsars
themselves should also be strange stars, but not normal neutron stars), we will mainly concentrate
on pulsar planets. So, as the initial step, we first select all the candidates of pulsar planets.
In this aspect, the EU data base contributes most of the objects. There are 18 pulsar planets listed in EU, 6
listed in ARCHIVE, and 3 listed in ORG. The total number of pulsar planet candidates is 19 after
considering the overlapping in different databases. In Table~1, we have listed some key parameters
of these 19 candidates as well as their host pulsars. Among these objects, 6 planets interestingly
have an orbital period less than 0.1 day (i.e., 8640 seconds). Note that our exact period criteria for
SQM objects is 6100 s, but we believe that all the planets with $P_{\rm orb} < 0.1$ days deserve being
paid special attention to.

The nature of companions around pulsars is actually not easy to be well defined. A companion
of several Jupiter mass could be a massive planet, but it could also be a small white dwarfs.
The key problem is that its radius usually could not be accurately measured. As a result, we should bear in
mind that the 19 objects listed in Table~1 are only candidates, not confirmed pulsar planets.
According to the confidence level, we have divided these 19 objects into three classes, the gold sample,
the silver sample, and the copper sample. In the gold sample, there are strong clues supporting the
objects as planets. In the silver sample, there are some clues hinting the objects as
planets \citep{Ray2017}. In the copper sample, the objects might be planets, but the evidence supporting the idea
is highly lacking. Interestingly, we find that three objects in the gold sample and one object in the
silver sample have periods less than 0.1 days. We will describe the details of these objects in the next
section.

SQM planets may also exist around white dwarfs, because these so called white dwarfs might actually
be strange quark dwarfs. So, we also select all the WD planets that have an orbital period less
than 0.1 day from EU. The total number of WD planets met such a requirement is 5, as listed in Table~2.

\begin{figure}
	\plotone{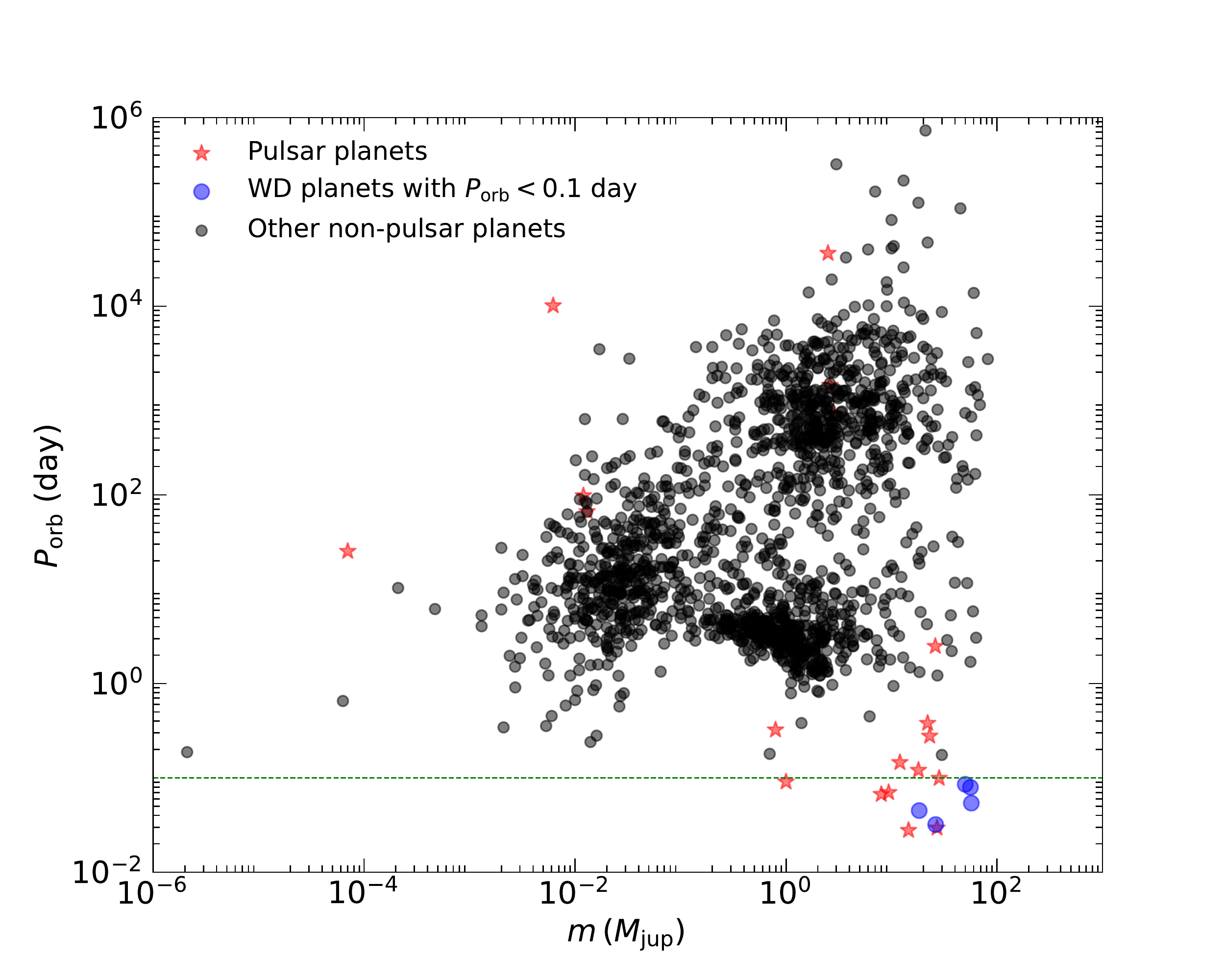}
	\caption{Orbital periods versus masses for all the 1638 exoplanets with data available from the EU
        web site (\url{http://www.exoplanet.eu/catalog/}). The red stars represent candidate pulsar planets.
        The blue points correspond to the five close-in WD planets with $ P_{\rm orb}<0.1 $ days, and
        the black dots represent other 1616 exoplanets.} \label{fig:fig1}
\end{figure}

To get an overall picture on how these short period planets differ from others, we have plot all the planets
with available masses and orbital periods on the $m$ -- $P_{\rm orb}$ plane in Figure~\ref{fig:fig1}.
It clearly shows that all the planets with a period smaller than 0.1 day are orbiting around
pulsars or white dwarfs. This kind of ultra-short period objects form a distinct group and
take a special place in the lower right region of Figure~\ref{fig:fig1}.
It strongly hints that they may have an exotic nature as compared with other planets.

\section{Candidates of SQM planets} \label{sec:close-in}

As explained in Section \ref{subsec:criteria}, because of the extreme compactness,
an SQM object could be very close to its host strange star, without being tidally disrupted.
So, closeness is a unique feature of SQM planets.
To search for SQM objects, we have selected all the close-in exoplanets around pulsars and WDs.
These ultra-short period (period less than 0.1 day) objects are listed in Table~\ref{tab:table3}.
To resist tidal disruption, they should have a relatively high mean density.
To see how exotic these objects are, we have calculated the minimum mean densities of these objects
by using the period-density relation of $ \rho_{\rm min}\approx{3\pi}/\left( {0.462^{3}G P_{\rm orb}^{2}}\right)
$ \citep{Frank1985,Bailes2011}. The results are also presented in Table~\ref{tab:table3}.
We can see that the minimum densities of these objects are
all significantly larger than that of normal rocky or iron material (typically with a density
of $\rm 1-10\,g\,cm^{-3} $). If these objects are planets but not small white dwarfs, then the
possibility that they are SQM objects is very high. Below, we will examine these close-in objects
one by one in detail and try to clarify their true nature.

\begin{deluxetable}{lccc}
	\tablecaption{Orbital parameters and minimum mean densities of ultra-short period objects \label{tab:table3}}
	\tablehead{
		\colhead{Planet name} & \colhead{$ P_{\rm orb} $} & \colhead{Orb. radius} & \colhead{$ \rho_{\rm min} $}\\
		\colhead{} & \colhead{(s)} & \colhead{$ a $ ($ 10^{10}\rm \,cm $)}  & \colhead{($\rm g \,cm^{-3} $)}
	}
	\startdata
	XTE J1807-294 b & 2404  & 3.1 & 247.9 \\
	XTE J1751-305 b & 2545  & 3.4 & 221.2 \\
	PSR 0636 b & 5789  & 5.4 & 42.8\\
	PSR J1807-2459A b & 6048  & 5.6 & 39.2 \\
	PSR 1719-14 b & 7837  & 6.6 & 23.3 \\
	PSR J2051-0827 b & 8563  & 7.1 & 19.5 \\
	GP Com b & 2765  & 2.1 & 187.5 \\
	V396 Hya b & 3888  & 2.6 & 94.8 \\
	J1433 b & 4666  & 4.0 & 65.8 \\
	WD 0137-349 b & 6863  & 4.1 & 30.4 \\
	SDSS J1411+2009 b & 7379  & 4.7 & 26.3 \\
	\enddata
\end{deluxetable}

\subsection{Close-in objects around pulsars}\label{subsec:identify}

The mass of planets can distribute in a very wide range. Some planets
can be very massive. In fact, the upper mass limit of planets has been
derived by many authors, which could be $ 43^{+14}_{-23}M_{\rm jup} $
\citep{Grether2006}, $ 42.5M_{\rm jup} $ \citep{Ma2014} and/or $ 60M_{\rm jup}$
\citep{Hatzes2015}. On the other hand, white dwarf cannot be too small and they
should have a lower mass limit. Recently, two low-mass white dwarf were reported, i.e.
SDSS J184037.78+642312.3 ($ 0.17M_\odot $) \citep{Hermes2012} and
SDSS J222859.93+362359.6 ($ 0.16M_\odot $) \citep{Hermes2013}. They hint that
white dwarfs maybe are unlikely to be less than 100 $M_{\rm jup}$.
In our Table~1, all the objects are significantly less massive
than the planetary mass limits, thus are reasonable candidates for planets.

Among all the close-in candidates in Table~\ref{tab:table3}, three are gold sample objects, one
is a silver sample object, and two are copper sample objects. The other five are WD planet candidates.
Here, we describe these objects one by one.

\subsubsection{Gold sample objects} \label{subsec:golden}

PSR J0636 b is a companion of the millisecond pulsar PSR J0636+5129 (spin period 2.87 ms) \citep{Stovall2014}.
It has a mass of $ 8 M_{\rm\,jup} $. Its orbital period is $ \sim 5789\rm \,s $, and
the orbital radius is correspondingly $\sim5.4\times10^{10} \rm \,cm $.
PSR J0636+5129 does not exhibit any eclipses caused by excess material in the system \citep{Stovall2014, Spiewak2016, Kaplan2018}.
PSR J0636 b is clearly identified as a planet by many authors \citep{Stovall2014, Spiewak2016, Spiewak2018}.
It is also explicitly listed as a planet by several planet databases, such as by EU, EXOKyoto,
PHLUPR (short for: Planetary Habitability Laboratory of the University of Puerto Rico at Arecibo\footnote{\url{http://phl.upr.edu/projects/habitable-exoplanets-catalog/top10}}),
GCEXO (short for: a General Catalogue of EXOplanets\footnote{\url{http://www.exoplaneet.info/index.html}}).

\smallskip
PSR J1807-2459A b is a companion of the millisecond pulsar PSR J1807-2459A (spin period
3.06 ms) \citep{Amico2001, Ransom2001,Lynch2012}.
This object has a mass of $ 9.4 M_{\rm\,jup} $, with an orbital period of $ \sim 6048\rm \,s $,
and correspondingly an orbital radius of $\sim5.6\times10^{10} \rm \,cm $. PSR J1807-2459A shows
no eclipses, but one can not rule out the possibility of eclipses at longer wavelengths \citep{Ransom2001, Lynch2012}.
PSR J1807-2459A b is identified as a planet by several authors\citep{Amico2001, Ray2017}.
Websites including this object in their planet catalogues are
EU, EXOKyoto, PHLUPR, and GCEXO.   

PSR 1719-14 b is a companion of the millisecond pulsar PSR J1719-1438 (spin period 5.7 ms) \citep{Bailes2011,Martin2016}.
It has a mass of $ 1 M_{\rm\,jup} $, with an orbital period of $ \sim 7837\rm \,s $,
and an orbital radius of $\sim6.6\times10^{10} \rm \,cm $.
It is identified as a planet by many researchers \citep{Bailes2011, Martin2016, Spiewak2018}.
Websites listing this object in their planet catalogues are EU, ARCHIVE, EXOKyoto, PHLUPR, and GCEXO.
PSR J1719-14 b was once considered to be a C/O dwarf in an ultra-compact low-mass
X-ray binary (UCLMXB) by \citet{Bailes2011}. However, since its mass is very low ($ 1 M_{\rm\,jup} $),
it is more likely to be a planetary object.  \citet{Horvath2012} explicitly argued that PSR J1719-14 b should be
an exotic strange object rather than a C/O dwarf. Very recently, \citet{Huang2017} also
identified PSR J1719-14 b as an ideal candidate of SQM planet.

\subsubsection{Silver sample objects}\label{subsec:silver}
PSR J2051-0827 b is a companion of the millisecond pulsar PSR J2051-0827 (spin period 4.5 ms) \citep{STAPPERS1996,Ray2017}.
It has a mass of $ 28.3 M_{\rm\,jup} $, with an orbital period of $ \sim 8563\rm \,s $,
and an orbital radius of $\sim7.1\times10^{10} \rm \,cm $. Its mass is within the planetary mass range.
While \cite{Ray2017} suggested this object as a planet, it has also been argued that it might be
a brown dwarf \citep{STAPPERS1996}. Websites including this object as a planet in catalogues
are EU, ARCHIVE, EXOKyoto, PHLUPR, and GCEXO. The orbital period and orbital radius of this object
are slightly larger than our strange planet criteria, but we suggest that it might be a good candidate
for SQM object and deserves paying special attention to.

\subsubsection{Copper sample objects}\label{subsec:copper}

XTE J1807-294 b is a companion of the millisecond X-ray pulsar XTE J1807-294 (spin period 5.25 ms) \citep{Markwardt2003a,Markwardt2003b,Campana2003,Kirsch2004,Falanga2005,Riggio2007,Patruno2010}.
It has a mass of $ 14.5\pm 8.5 M_{\rm\,jup} $, with an orbital period of $ \sim 2404\rm \,s $,
and an orbital radius of $\sim3.1\times10^{10} \rm \,cm $.
No X-ray eclipse was observed from this system \citep{Falanga2005}.
According to the mass-radius relation, the companion may be the core of a
previously crystallized C/O dwarf \citep{Deloye2003}.
However, there are no emission or absorption lines found from this companion \citep{Campana2003}.
This object is listed as a planet in EU and EXOKyoto, but its true nature is still highly unclear.

XTE J1751-305 b is a companion of the millisecond X-ray pulsar XTE J1751-305 (spin period 2.3 ms)
\citep{Markwardt2002,Gierli2005, Andersson2014}.
It has a mass of $ 27\pm 10 M_{\rm\,jup} $, with an orbital period of
$ \sim 2545\rm \,s $, and an orbital radius of $\sim3.4\times10^{10} \rm \,cm $.
The pulsar show no X-ray eclipses during observations \citep{Markwardt2002,Gierli2005}.
This object is listed as a planet in EU and EXOKyoto, but several authors have also
argued that it may be the core of a previously crystallized C/O dwarf \citep{Deloye2003}.

The orbital parameters of both XTE J1807-294 b and XTE J1751-305 b well satisfy
our SQM criteria. Also, it is obvious that the masses of both objects are within the planet mass range.
Although their true nature is still uncertain, we interestingly notice that \citet{Horvath2012} have
argued that an exotic strange object interpretation is the best alternative to a C/O dwarf
interpretation for these two objects. We believe that the likelihood of these two objects
being SQM planets is high. They need to be studied in more detail in the future.

\subsection{Close-in objects around white dwarfs}\label{subsec:close-in WD}

GP Com b is a companion of the white dwarf GP Com \citep{Nather1981,Kupfer2016}.
Its mass is $ 26.2\pm16.6 M_{\rm\,jup}$, with an orbital period of
$ \sim 2765\rm \,s $, and an orbital radius of $\sim2.1\times10^{10} \rm \,cm $ \citep{Kupfer2016}.
There are suggestions that this object may be a degenerated He dwarf \citep{Nather1981}.
But the observed abundances of Ne line from this object could be affected by crystallization processes in the core,
and this excludes the highly evolved He donor nature for it \citep{Kupfer2016}.
Alternatively, it was argued to be a planet by many authors \citep{Cunha2018, Wong2018}. Websites listing this
object as a planet are EU, EXOKyoto, PHLUPR, and GCEXO.

V396 Hya b is a companion of the white dwarf V396 Hya  \citep{RUIZ2001,Kupfer2016}.
The orbital period of the planet is $ \sim 3888\rm \,s $, and the orbital radius
is $\sim 2.6\times10^{10}\rm\, cm $.  Its mass is measured as $ 18.3\pm12.2 M_{\rm\,jup} $ \citep{Kupfer2016}.
It was suggested to be a degenerated He dwarf \citep{RUIZ2001}, or a crystallized Ne core  \citep{Kupfer2016}.
However, other authors \citep{Cunha2018, Wong2018} have argued that it could be a planet.
Websites listing this object as a planet are EU, EXOKyoto, and PHLUPR.

J1433 b is a companion of the white dwarf SDSS J143317.78+101123.3 (WD J1433)
\citep{Littlefair2006, Santisteban2016}. It has an orbital period of
$ \sim 4666\rm \,s $, and an orbital radius of $\sim 4.0\times10^{10} \rm \,cm $.
Its mass is $ 57\pm0.7 M_{\rm\,jup} $. It was argued to be a planet by several
research groups \citep{Cunha2018, Wong2018}. Websites listing this object as a planet
are EU, EXOKyoto, and PHLUPR. However, note that a few other authors suggested that
this object may be an irradiated brown dwarf \citep{Santisteban2016}.

WD 0137-349 b is a companion of the white dwarf 0137-349
\citep{Maxted2006, Burleigh2006, Littlefair2014, Casewell2015, Longstaff2017}.
It has a mass of $ 56\pm6 M_{\rm\,jup} $, with an orbital period of $ \sim 6863\rm \,s $,
and an orbital radius of $\sim 4.1\times10^{10} \rm \,cm $. This object is listed as
a planet in EU and EXOKyoto databases. But again note that it was suggested to
be an irradiated brown dwarf by several authors \citep{Maxted2006, Burleigh2006}.

SDSS J1411+2009 b is a companion of the white dwarf SDSS J141126.20+200911.1 (WD J1411).
It has a mass of $ 50\pm2.0 M_{\rm\,jup}$, with an orbital period of $ \sim 7379\rm \,s $ and
an orbital radius of $ \sim 4.7\times10^{10} \rm \,cm $ \citep{DRAKE2010,Beuermann2013,Littlefair2014}.
It is listed as a planet in EU and EXOKyoto databases, but several authors have also
suggested it as an irradiated brown dwarf \citep{Beuermann2013}.

Orbital parameters of the above five close-in objects around white dwarfs satisfy the
criteria of $P_{\rm orb}<0.1 $ day. Their masses are also within the planetary mass range.
However, the planetary nature of these objects is still debatable. Especially,
they may actually be brown dwarfs. Here, we give some more discussion on this point.
In fact, there is no clear boundary between the masses of planets and brown dwarfs.
It is well known that the mass of brown dwarfs can range from the
Deuterium-burning limit ($ 0.013M_\odot (\sim 13 M_{\rm jup})$) to the
Hydrogen-burning limit ($ 0.072M_\odot (\sim 75 M_{\rm jup})$).
The property of a close-in companion is usually seriously affected by the
irradiation from its host since it is generally tidally
locked \citep{Demory2011, Laughlin2011, Burgasser2019}. This effect is
quite similar for both brown dwarfs and giant planets, thus could not be
easily used to discriminate them \citep{Faherty2013}.
However, we notice that three of the five objects have extremely small orbital periods.
They are GP Com b, V396 Hya b, and J1433 b, and their orbital periods are 2765 s,
3888 s, and 4666 s. As a result, their minimal possible mean density is 187.5 $\rm g/cm^3$,
94.8 $\rm g/cm^3$, and 65.8 $\rm g/cm^3$, respectively. The density is so high that they can
hardly be normal brown dwarfs. We argue that at least these three objects are very good
candidates for SQM planets.

\section{GW from SQM planetary systems} \label{sec:GW}

According to general relativity, a binary system continuously emits GW signals due to
the orbital motion of the companion. It will lead to an evolution of the orbit,
and make the GW emission power increase gradually.
At some stage of this gradual process,  GW detectors such as LISA
(Laser Interferometer Space Antenna) will be able to detect
the GW signals from these systems \citep{Cunha2018, Wong2018}.

For close-in companions orbiting around their hosts, GW emission may be a powerful tool
to probe their nature. In this section, we first calculate the persistent GW emissions
from the candidate SQM systems in our sample and evaluate the possibility of being detected by
the LISA observatory. Then, we also calculate the
strength of the catastrophic GW bursts when the candidate SQM systems
finally merge due to continuous GW emissions, and compare the results with relevant
GW detectors.

\subsection{Persistent GWs from SQM planet systems}

\begin{figure}
	\plotone{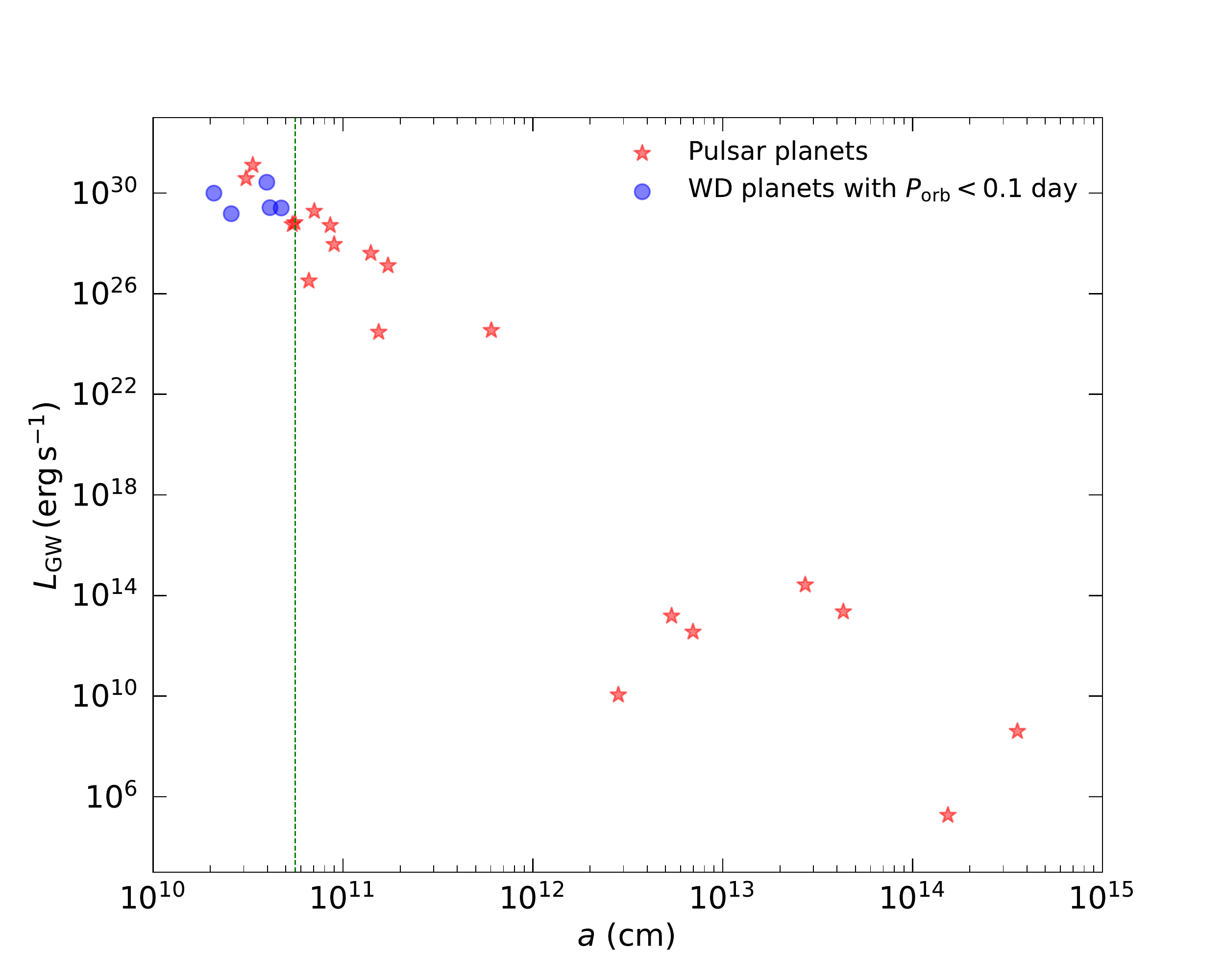}
	\caption{Power of gravitational wave emission versus orbital radius for the planetary systems
        of our sample. The red stars are pulsar planets and the blue points are WD planets
		with $ P_{\rm orb}<0.1 $ day. The vertical green dashed line markes the critical
		tidal disruption radius of $ a=5.6\times10^{10}\rm\,cm $ for normal matter planets. \label{fig:fig3}}
\end{figure}

\begin{figure}
	\plotone{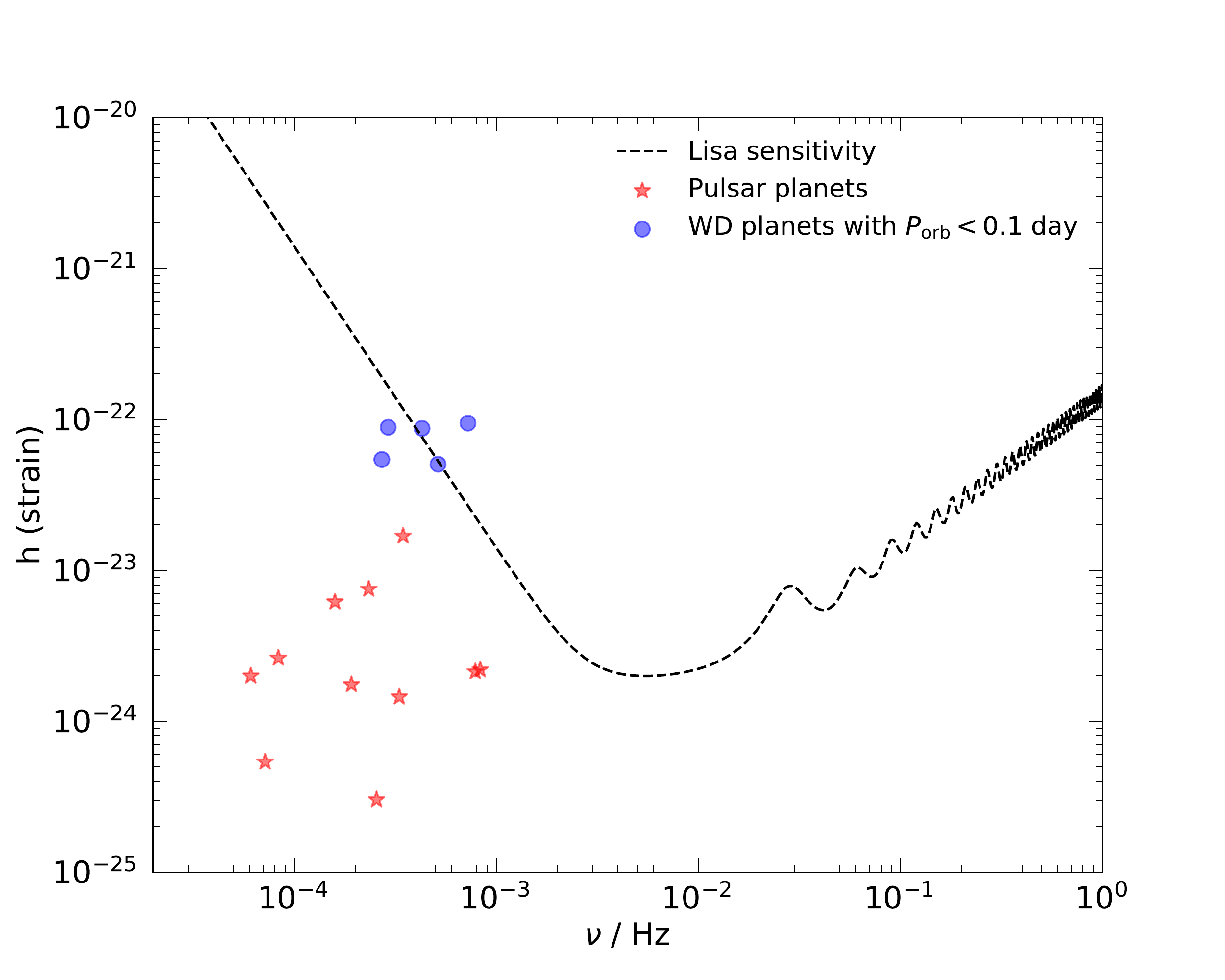}
	\caption{GW strain amplitude versus frequency for the planetary systems of our sample.
		The blue points represent WD planets with $ P_{\rm orb}<0.1 $
		day and the red stars represent pulsar planets.
		The black dashed line represents the sensitivity curve of LISA with an one year integration time.
		For a similar plot, please also see \citet{Cunha2018} and \citet{Wong2018}. \label{fig:fig4}}
\end{figure}

For all the planetary systems of our sample, we have calculated their persistent
GW luminosity and GW strain amplitude. The results are presented in Table~\ref{tab:table4}.
In Figure~\ref{fig:fig3}, we plot the GW luminosity versus the orbital radius
for them. The red stars and blue points represent pulsar planets and WD planets,
respectively. Generally speaking, the orbital radius is a key parameter determining
the GW power. For those systems with the orbital radius being less than the critical
tidal disruption radius of $5.6\times10^{10}\rm\,cm$, the GW power is much stronger.
Thus there is a hope that GW emission from these systems could be detected by our GW
detectors.

In Figure~\ref{fig:fig4}, we plot the GW strain amplitude against GW frequency
for the planetary systems of our sample. In this plot, the red stars represent pulsar planets
and the blue points represent WD planets. The black line is the one-year integration
sensitivity curve of LISA \footnote{\url{http://www.srl.caltech.edu/~shane/sensitivity/index.html}}.
Figure~\ref{fig:fig4} shows clearly that three ultra-short period systems
(GP Com b, V396 Hya b and J1433 b) are lying above the sensitivity curve of LISA, thus
may hopefully be detected by this powerful GW observatory. The two very close-in systems
containing XTE J1807-294 b and XTE J1751-305 b are below the sensitivity curve, since their distances
are still too large. If these two systems were located within a distance of $ 400\rm\,pc $,
then they would be detectable to LISA.
For close-in planetary systems, GW observation can provide key information
on the planet mass, the orbital period, and the orbital radius. We argue that
GW observation would be a unique tool to search for SQM candidates. In the future,
if a very close-in planet-like object (with the mass being in the planet range, and the
orbital period significantly less than 6100 s) could be found orbiting around a stellar
object through GW observations, then it must be an SQM planetary system.

\begin{deluxetable}{lccc}
	\tablecaption{GW Luminosity, strain amplitude, and coalescence time scale for the
planetary systems of our sample. \label{tab:table4}}
	\tablehead{
		\colhead{Planet name} & \colhead{$ L_{\rm GW} $} & \colhead{$\rm h $} & \colhead{$ t_{\rm co} $}\\
		\colhead{} & \colhead{($ \rm egr\,s^{-1} $)} & \colhead{}  & \colhead{(yr)}
	}
	\startdata
	XTE J1807-294 b & $ 3.8\times10^{30} $ & $ 2.2\times10^{-24} $ & $ 1.9\times10^{8} $ \\
	XTE J1751-305 b & $ 1.3\times10^{31} $ & $ 2.1\times10^{-24} $ & $ 1.1\times10^{8} $ \\
	PSR 0636 b & $ 5.7\times10^{28} $ & $ 1.7\times10^{-23} $ & $ 3.7\times10^{9} $ \\
	PSR J1807-2459A b & $ 6.7\times10^{28} $ & $ 1.5\times10^{-24} $ & $ 3.5\times10^{9} $ \\
	PSR 1719-14 b & $ 3.2\times10^{26} $ & $ 3.0\times10^{-25} $ & $ 6.6\times10^{10} $ \\
	PSR J2051-0827 b & $ 1.9\times10^{29} $ & $ 7.5\times10^{-24} $ & $ 3.0\times10^{9} $ \\
	PSR J1544+4937 b & $ 5.2\times10^{28} $ & $ 1.7\times10^{-24} $ & $ 6.9\times10^{9} $ \\
	PSR J2241-5236 b & $ 9.1\times10^{27} $ & $ 6.2\times10^{-24} $ & $ 2.0\times10^{10} $ \\
	PSR J1446-4701 b & $ 4.1\times10^{27} $ & $ 2.6\times10^{-24} $ & $ 5.7\times10^{10} $ \\
	PSR J2322-2650 b  & $ 3.0\times10^{24} $ & $ 5.4\times10^{-25} $ & $ 2.4\times10^{12} $ \\
	PSR B1957+20 b & $ 1.3\times10^{27} $ & $ 2.0\times10^{-24} $ & $ 1.4\times10^{11} $ \\
	PSR J1502-6752 b & $ 3.5\times10^{24} $ & $ 2.5\times10^{-25} $ & $ 1.7\times10^{13} $ \\
	PSR 1257 12 b & $ 1.1\times10^{10} $ & $ 8.4\times10^{-31} $ & $ 3.1\times10^{21} $ \\
	PSR 1257 12 c & $ 1.5\times10^{13} $ &  $ 8.2\times10^{-29} $ & $ 2.2\times10^{20} $ \\
	PSR 1257 12 d & $ 3.6\times10^{12} $ & $ 5.8\times10^{-29} $ & $ 6.7\times10^{20} $ \\
	PSR B0943+10 b & $ 2.7\times10^{14} $ & $ 3.0\times10^{-27} $ & $ 5.8\times10^{20} $ \\
	PSR B0943+10 c & $ 2.3\times10^{13} $ & $ 1.7\times10^{-27} $ & $ 4.0\times10^{21} $ \\
	PSR B0329+54 b & $ 1.8\times10^{5} $ & $ 9.7\times10^{-31} $ & $ 3.1\times10^{26} $ \\
	PSR B1620-26(AB) b & $ 4.0\times10^{8} $ & $ 4.3\times10^{-29} $ & $ 2.4\times10^{25} $ \\
	GP Com b & $ 9.9\times10^{29} $ & $ 9.5\times10^{-23} $ & $ 4.2\times10^{8} $ \\
	V396 Hya b & $ 1.5\times10^{29} $ & $ 5.1\times10^{-23} $ & $ 1.5\times10^{9} $ \\
	J1433 b & $ 2.7\times10^{30} $ & $ 8.7\times10^{-23} $ & $ 4.3\times10^{8} $ \\
	WD 0137-349 b & $ 2.6\times10^{29} $ & $ 8.9\times10^{-23} $ & $ 2.0\times10^{9} $ \\
	SDSS J1411+2009 b & $ 2.6\times10^{29} $ & $ 5.4\times10^{-23} $ & $ 2.2\times10^{9} $ \\	
	\enddata
\end{deluxetable}

\subsection{GW bursts from merging SQM planet systems}
\begin{figure}
	\plotone{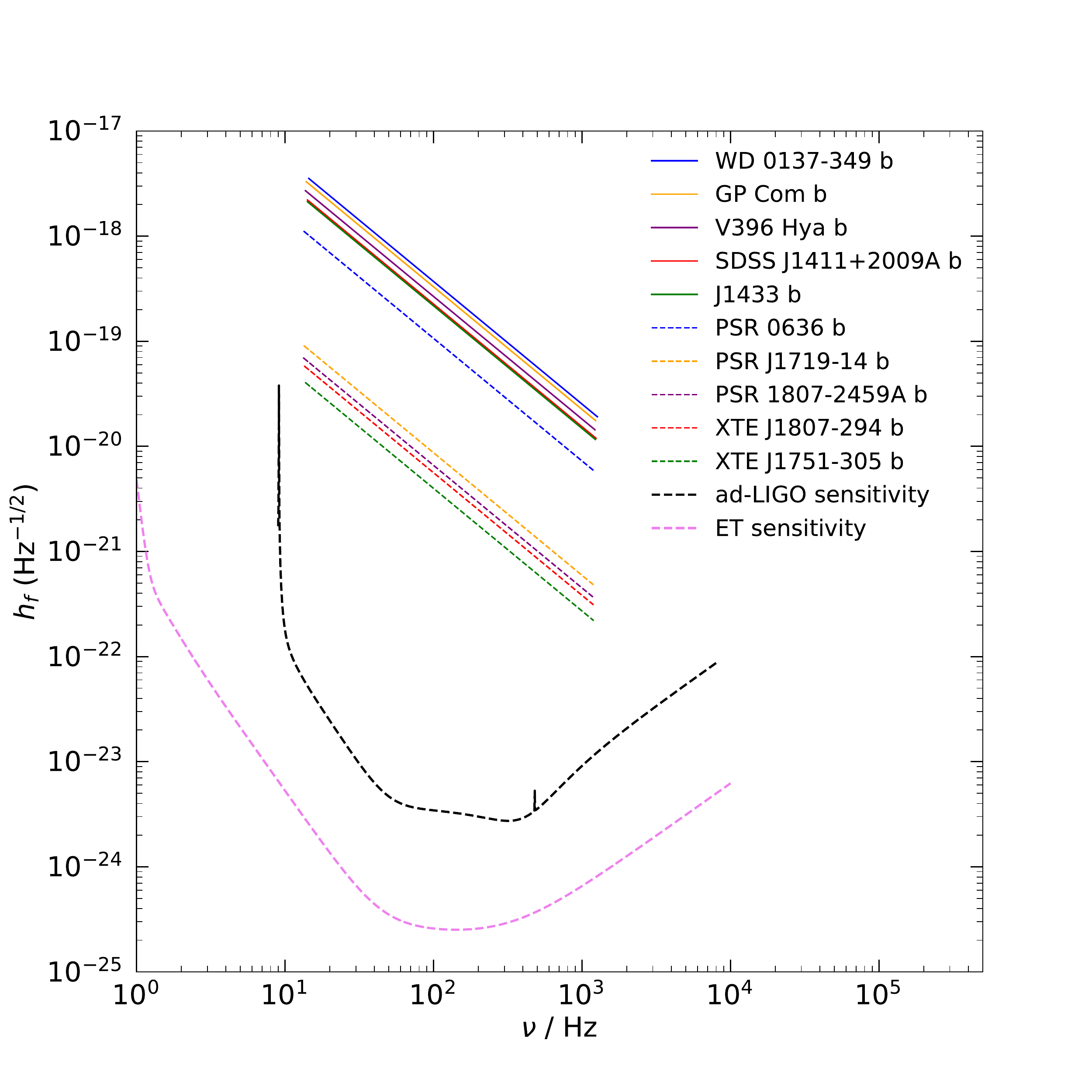}
	\caption{Strain spectral amplitude of the GW bursts for
		coalescing SS and SQM planet systems. The sensitivity curves of
        the advanced LIGO and Einstein Telescope are also plotted. \label{fig:fig5}}
\end{figure}

Due to the self-gravity and strong self-bound force of strange quark matter,
an SQM planet can get very close to its host without being tidally disrupted by tidal force.
During the spiral-in process, the separation between the two objects decreases with
time until they merge with each other. At the final merging stage,
the system will give birth to a strong GW burst \citep{Geng2015}. In Figure~\ref{fig:fig5},
we have plot the strain spectral amplitudes of the GW bursts that will be produced
by several candidate SQM planetary systems. Note that these systems are at different
distances, and the planets have different masses. For each system, we have used the
actually observed parameters in the calculation. We see that the GW amplitudes are all well
above the sensitivity curves of both the advanced LIGO and the Einstein Telescope, thus
can potentially be detected by these instruments.

\begin{figure}
	\plotone{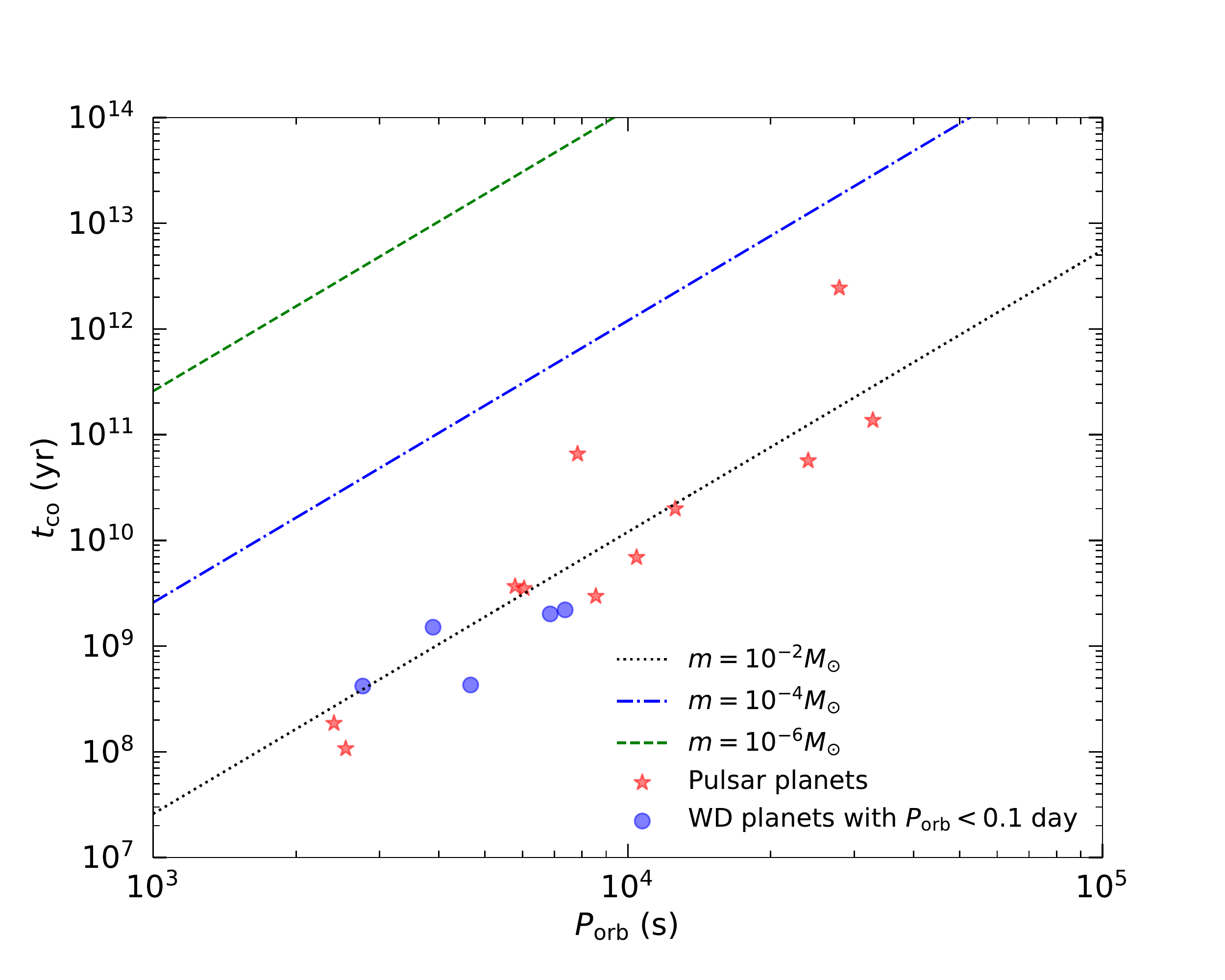}
	\caption{Coalescence timescale versus the initial orbital period for the candidate SQM
        planetary systems in our sample. Red stars correspond to pulsar planets and blue
        points correspond to white dwarf planets. The three straight lines illustrate the
        coalescence timescale for three different planet masses, with the host mass being
        set as $1.4 M_\odot$. \label{fig:fig6}}
\end{figure}

The energy loss rate due to GW emission is generally small as compared with the
total kinetic energy of the planet. It may take a long time for a planetary system
to finally merge. The merger timescale is mainly determined by the orbital radius
and the planet mass. In Table~\ref{tab:table4}, we have also calculated the coalescence
timescales of the planetary systems in our sample. The results are illustrated in
Figure~\ref{fig:fig6}. While most systems essentially will not be able to merge
even in the lifetime of the Universe, there are about 10 close-in systems that
would interestingly merge on a timescale of $10^8$ --- $10^9$ yr.
For example, the merger timescale is $ \sim10^{8} $yr for the planetary systems
of XTE J1807-294, XTE J1751-305, GP Com b, and J1433 b.
Additionally, other factors may be involved and may lead to a much rapid merging
process. For example, a pulsar may have multiple planets and the complicate
interaction between these companions may speedup the merging processes of some
objects \citep{Huang2014}.

In short, merging of an SQM planet with its host pulsar can essentially happen on an
expectable timescale in our Galaxy. GW emission from these events can be well detected
by our current and future detectors.
We suggest that searching for GW signals from merging planetary systems could be
set as an important goal for advanced LIGO and Einstein Telescope. It deserves extensive
efforts since it can provide a unique test for the SQM hypothesis.

\section{Conclusions} \label{sec:conclusion}

In this study, we have tried to search for SQM planet candidates among extra-solar planetary systems.
The criteria for SQM planets is set as $a <5.6\times10^{10} \rm \, cm $ and/or $P_{\rm orb} < \rm 6100\, s $.
A planet lying closer than this limit with respect to its host will need to have a density significantly
larger than 30~g~cm$^{-3}$ to resist the tidal force, thus is unlikely a normal matter planet,
but should be an SQM object. As a result, we find that 11 objects are good candidates for SQM planets,
including 3 gold sample objects, 1 silver sample object, 2 cooper sample objects, and 5 white dwarf
companions. The three gold sample objects are PSR 0636 b, PSR J1807-2459A b, and PSR J1719-14 b. Their
masses are all less than 10 $M_{jup}$ and their possibility of being a planetary object is very high.
Among them, although PSR 1719-14 b has a period (7837 s) slightly larger than 6100 s, we still list it
as a good candidate since it is essentially in a very close-in orbit. The silver sample object
(PSR J2051-0827 b), the two cooper sample objects (XTE J1807-294 b, XTE J1751-305b), and the five
white dwarf companions (GP Com b, V396 Hya b, J1433 b, WD 0137-349 b, SDSS J1411+2009 b) are all interesting
candidates, but whether they are planetary objects or white dwarfs is still highly uncertain and
need further clarification.
We have also calculated the GW emissions from these systems. It is found that persistent GW emissions
from at least three of them are detectable to LIGO even on a one-year integration. More encouragingly,
GW bursts produced at the final merging stage by these candidate SQM planets are well above the
sensitivity curves of advanced LIGO and Einstein Telescope. GW observations thus could
be a promising strategy for testing the SQM hypothesis.

It is striking to note that our SQM candidates are mainly found around millisecond pulsars.
It leads to the interesting conjecture that there might be some intrinsic connection between SQM objects and
low mass X-ray binaries (LMXBs).
Indeed, some authors \citep{Li1995,Xu2002,Xu1998,Poutanen2003,Zhu2013} have tried to identify
SSs in LMXBs. For example, the famous LMXBs of Her X-1 \citep{Li1995} and
SAX J1808.4-3658 have been argued as SS candidates \citep{Li1999,Poutanen2003,Gangopadhyay2012}.
\citet{Poutanen2003} and \citet{Gangopadhyay2012} also noticed the similarity of
XTE J1807-294 and XTE J1751-305 with respect to SAX J1808.4-3658 when they argued that
SAX J1808.4-3658 should be a strange star.
Furthermore, \citet{Gangopadhyay2013} listed 12 stars in binary systems as SSs, again
including Her X-1 and SAX 1808.4-3658.
Recently, \citet{Chen2016} pointed out that the binary systems of SAX 1808.4-3658 and PSR J1719-1438
may have similar evolutionary history.
In fact, the link between strange stars and LXMBs is not difficult to understand theoretically.
Continuous accretion and significant mass transfer widely exists in LXMBs. Increase of the mass
can easily lead to an ultra-high density at the center of the pulsar, leading to a phase transition
and turn the pulsar into a strange quark star even it is originally born as a neutron star.

Pulsars in these close-in binary systems generally show no eclipsing in high-frequency range.
There are two possible reasons for this. First, the inclination angle of the orbit should be
relatively large. Second, the density of the companion may be high and its radius is correspondingly
very small. This will further support the SQM nature of the object. In several cases, possible eclipse
is reported to be observed at low-frequency range. The small amount of eclipsing plasma
in these cases may come from the ablation of the outer crust of the SQM planet.

\section{Acknowledgments}
This work was supported by the National Natural Science Foundation of China (Grant
Nos. 11873030, 11475085, 11535005, 11690030, and 11703041), by Nation Major State Basic Research and
Development of China (2016YFE0129300), and by the Strategic Priority Research Program of
the Chinese Academy of Sciences (``multi-waveband Gravitational Wave Universe'', Grant
No. XDB23040000).






\nocite{*}
\bibliographystyle{aasjournal}%
\bibliography{apjbib}%



\end{document}